%% file: main.tex
\documentclass[acmsmall,screen,nonacm]{acmart}
\title{Goanna: Resolving Haskell Type Errors With Minimal Correction Subsets}
\input{preamble.tex}

\begin{document}
\begin{abstract}

Statically typed languages offer significant advantages, such as bug prevention, enhanced code quality, and reduced maintenance costs. However, these benefits often come at the expense of a steep learning curve and a slower development pace. Haskell, known for its expressive and strict type system, poses challenges for inexperienced programmers in learning and using its type system, especially in debugging type errors. We introduce Goanna, a novel tool that serves as a type checker and an interactive type error debugging tool for Haskell. When encountering type errors, Goanna identifies a comprehensive list of potential causes and resolutions based on the minimum correction subsets (MCS) enumeration. We evaluated Goanna's effectiveness using 86 diverse Haskell programs from online discourse, demonstrating its ability to accurately identify and resolve type errors. Additionally, we present a collection of techniques and heuristics to enhance Goanna's suggestion-based error diagnosis and show their effectiveness from our evaluation.
\end{abstract}
\maketitle

\section{Introduction} \label{sec:introduction}
    
Statically typed languages have gained popularity in the mainstream programming world \cite{StackOverflow2022-aw}. Many new languages have been designed with strict type systems, while others have introduced static typing through external tools. Numerous studies indicate that programming with strongly typed languages can prevent certain errors \cite{Bogner2022-vf}, improve code quality \cite{Mayer2012-lg}, and reduce maintenance costs \cite{Kleinschmager2012-bg} compared to similarly positioned dynamic languages \cite{Bogner2022-vf}. Despite their increasing popularity and benefits, challenges persist in the real-world adoption of these languages \cite{Zeng2019-ou}. The steep learning curve of complex type systems remains an obstacle to their adoption. 

Haskell is renowned for its expressive and robust type system. It enables programmers to model complex problems as constructs and relations within type systems and to develop robust programs in a type-driven style. Historically, many type system innovations initially introduced by Haskell \cite{Hudak2007-kn}, including algebraic data types, type inference, and type classes, have now found their way into mainstream programming languages \cite{TypeScriptTeam_undated-qk,Klabnik_undated-mp,Griesemer_undated-ff}.
    
\begin{figure}
        \centering
        \includegraphics[width=\linewidth]{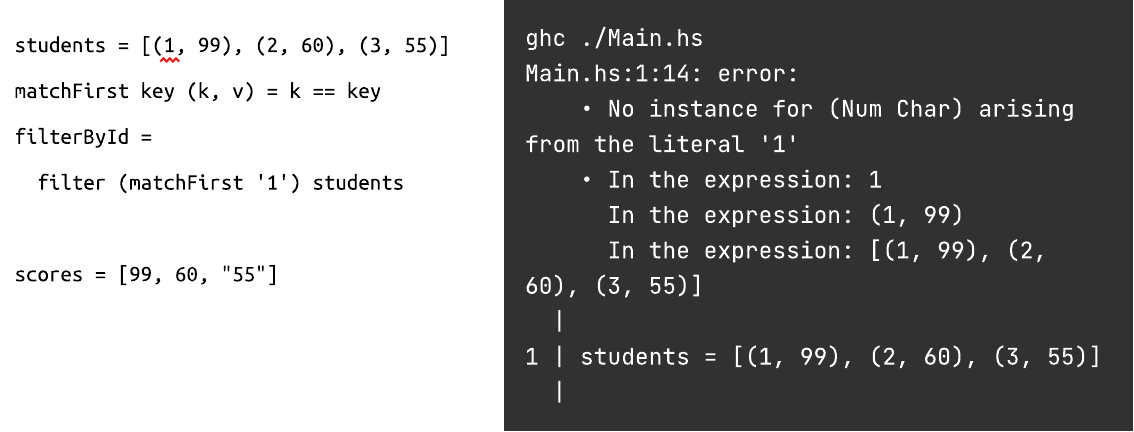}
        \caption{Inspecting a type error using the Haskell compiler GHC (Glasgow Haskell Compiler)}
        \label{fig:motivation}
    \end{figure}

However, Haskell is also known for its steep learning curve and unforgiving type errors. Numerous research efforts have attempted to address these challenges \cite{Tirronen2015-nr,Chen2014-dz, Heeren2003-kd,Zhang2015-xy, Lerner2007-mu,Zhang2017-tj}. The type errors generated by the most commonly used Haskell compiler, GHC (Glasgow Haskell Compiler), often lead to confusion among novice users, and sometimes experts. For instance, in the program shown in Fig.~\ref{fig:motivation}, a type mismatch between a {\tt Char} type and an integer number type results in a perplexing type error for novice users. We have identified three challenges to making use of these error messages.

    \begin{enumerate}
        \item They fixate on one possible cause, while other potential causes exist.
        \item Changing the suggested location may not completely rectify the error.
        \item Not enough contextual information is given for programmers to understand how the judgment was made.

    \end{enumerate}

    To address these challenges of diagnosing and fixing type errors in Haskell, we present a new tool: \textit{Goanna}. Goanna is a Haskell type-checker based on Minimal Correct Set (MCS) enumeration. Compared to traditional type-checking tools, Goanna provides improved type error reporting by providing a comprehensive list of possible causes and suggesting valid fixes for each cause.  Goanna differs from the past type debugging systems (as reviewed in Section \ref{sec:related-work}) through its use of Minimal Correction Subsets (MCS), where a single MCS represents a complete set of locations that constitutes a possible cause.

	To further enhance Goanna's support for type-error resolution, we provide optimization strategies (Section~\ref{sub:optimization}) to identify and reduce the unhelpful suggestions, as well as ranking heuristics (Section~\ref{sub:ranking}) to suggest more likely fixes first. Additionally, we provide Goanna-IDE, an interactive debugging front-end designed to efficiently navigate and interpret Goanna's type error diagnosis.

    We conducted empirical studies that evaluated Goanna's accuracy (Section \ref{sub:eval-accuracy}), conciseness (Section \ref{sub:eval-conciseness}) and performance (Section \ref{sub:eval-performacne}). Our evaluation shows that, compared to other type-checking tools, Goanna consistently provides accurate error diagnostics and correct fixes in its top suggestions. We also demonstrate that Goanna generally offers a concise list of possible causes, thanks to its cause optimization process. Although Goanna may not consistently provide instantaneous results for real-time feedback, it can deliver on-demand diagnoses when programmers require additional assistance.

    The key contributions of this research include: \begin{itemize}
        \item Goanna, an open-sourced Haskell type checker with improved error reporting based on MCS enumeration and program slicing;
        \item Goanna-IDE, an open-sourced interactive type error debugging interface for Haskell; 
        \item A collection of heuristics and optimization techniques to enhance MCS-based type error reporting; and
        \item An evaluation of Goanna's accuracy, conciseness, and performance.
    \end{itemize}

  The techniques we used in Goanna, such as MCS enumeration and heuristics for ranking possible causes, are not exclusive to Haskell. Rather, they apply to statically typed programming languages, in general. We intentionally designed Goanna to use a modular architecture that can be easily extended to support other programming languages with similar typing disciplines.

    \section{Goanna-IDE Walkthrough} \label{walkthrough}
    We first illustrate Goanna's capability by demonstrating the usage of Goanna-IDE. Goanna-IDE is a type error debugging interface for Haskell. It is designed to efficiently navigate and make use of Goanna's type error diagnosis through visualization and interactivity. Goanna-IDE provides comprehensive diagnostic error messages for type errors in Haskell and allows programmers to interactively explore their options. An online demo of Goanna-IDE is available for evaluation~\cite{Anonymous2023-bo}. Goanna-IDE includes a file explorer, a text editor, and a debugging panel. Goanna-IDE provides the following features when type errors are encountered:

    \begin{itemize}
        \item Thoroughly detect all type errors within the codebase and allow users to inspect each type error individually via the debugging panel.
        \item Indicate the most likely causes by star indicators.
        \item Show necessary type hints in the editor panel to help reason about each possible cause.
        \item Allow users to trace type errors across multiple files.
    \end{itemize}

    \subsection{Examples of Diagnosing Type Errors with Goanna-IDE}

\begin{figure}[htb!]
        \centering
        \includegraphics[width=\linewidth]{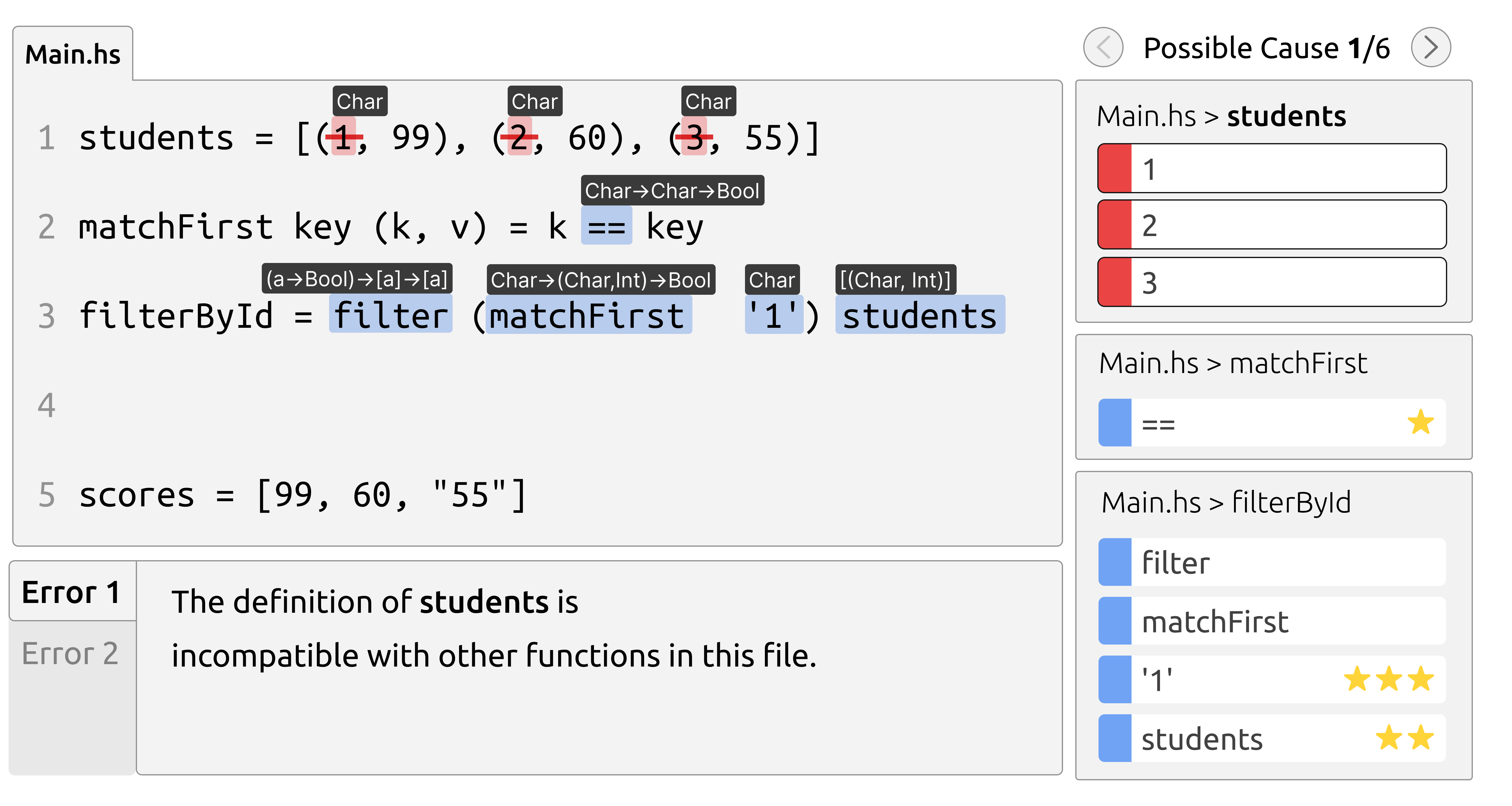}
        \caption{\textbf{Goanna's error diagnosis} Goanna shows that to fix the type error, the literals \texttt{1}, \texttt{2}, and \texttt{3} on line 1 need to be changed to Char type.}
        \label{fig:goanna-example-1}
    \end{figure}

    For the type error in the motivating example, Goanna shows 6 possible causes of the error (see the upper right corner of Fig.~\ref{fig:goanna-example-1}). When focusing on the cause suggested by GHC, instead of highlighting only the literal \texttt{1}, Goanna reports all 3 literals that needed to be changed all at once if the programmer chooses to address this cause. In addition, Goanna also indicates that these integer literals need to be changed to \texttt{Char} type using the inlay type hints on line 1, largely narrowing down the potential ideal fixes.

    Note that the cause suggested by GHC is only one of the possibilities identified by Goanna. In fact, Goanna suggests that there are more likely fixes, indicated by the star symbols. The most likely fix, based on Goanna's cause heuristics (Section \ref{sub:ranking}), is the \texttt{Char} literal \texttt{'1'} on line 3, indicated by the 3 stars (Fig.~\ref{fig:goanna-example-1}).

\begin{figure}[htb!]
        \centering
        \includegraphics[width=\linewidth]{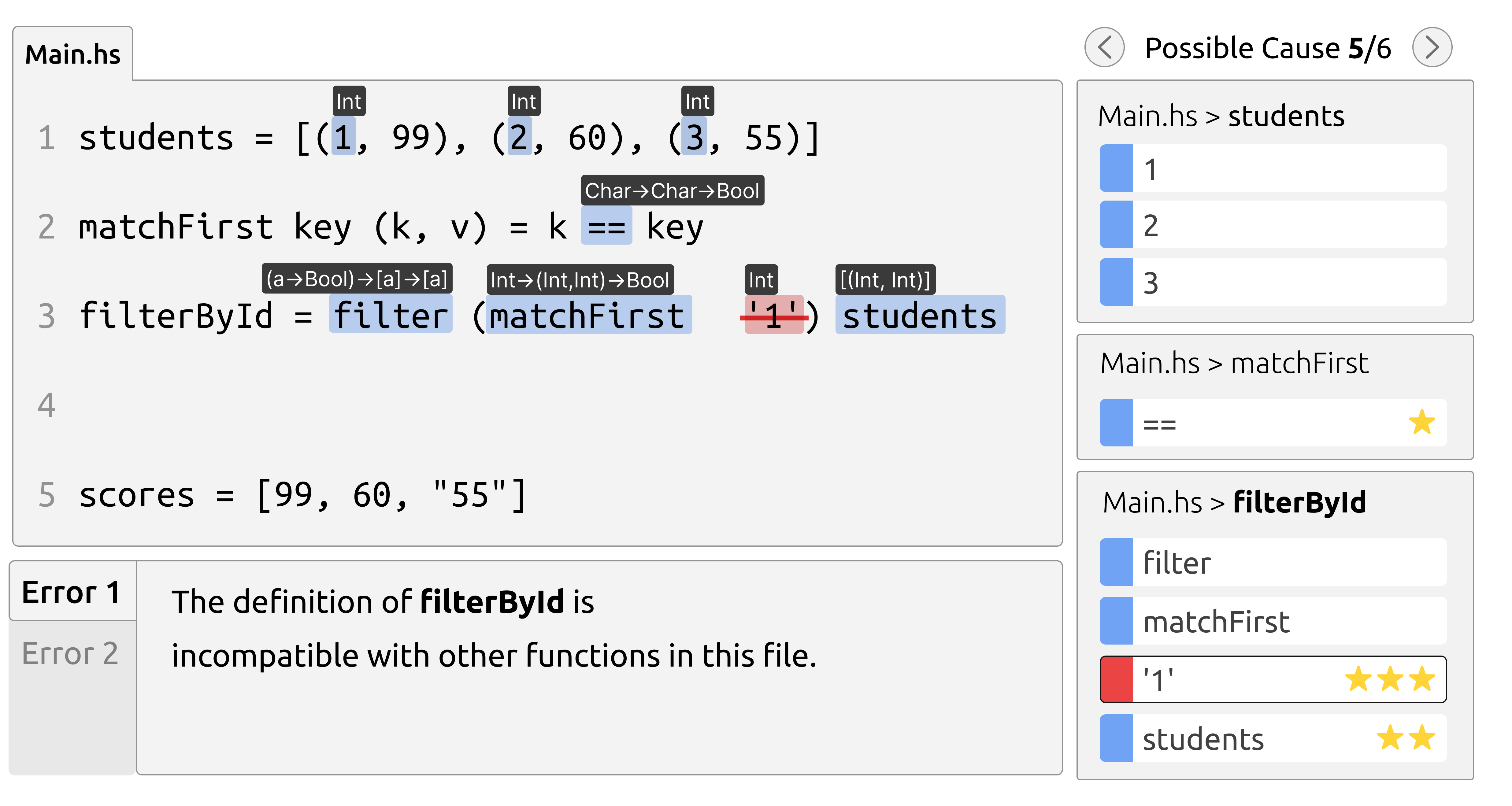}
        \caption{\textbf{Goanna's error diagnosis.} Goanna shows that the type error can be fixed by changing the literal $'1'$ on line 3, which needs an \texttt{Int} type. This, according to Goanna, is the most likely cause of the type error.}
        \label{fig:goanna-example-2}
    \end{figure}

    By clicking on the most likely cause, Goanna shows different highlights in the editor (e.g., see Fig.~\ref{fig:goanna-example-2}). Goanna reports that the error is caused by the literal \texttt{'1'} and suggests changing to an integer. All the type hints are adjusted based on our new assumption. Goanna ranks all possible causes using a series of heuristics. In this case, the preference is largely influenced by how many locations are required to be changed to fix the error. 
    
    \subsection{Identifying all type errors} \label{sub:all-errors}
    
\begin{figure}[htb!]
        \centering
        \includegraphics[width=\linewidth]{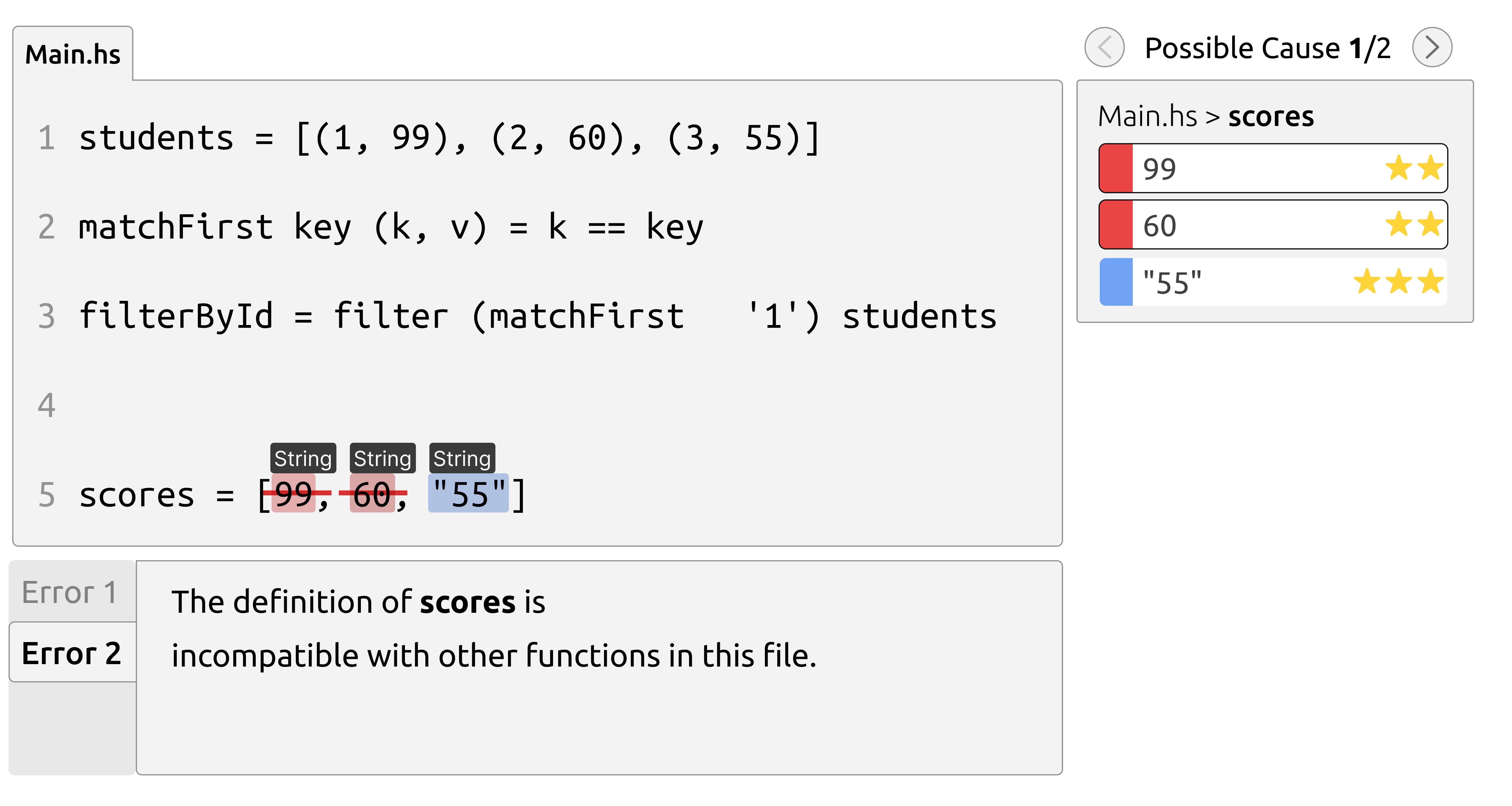}
        \caption{\textbf{Selecting a different error in Goanna.} Selecting a different error using Goanna's error selector. The debugging panel will show potential cause locations for the selected error. The highlights and type hints in the editor panel will focus on the selected error.}
        \label{fig:multi-error}
    \end{figure}

    A key feature of Goanna is its ability to detect all type errors in the code thanks to its MCS enumeration (Subsection \ref{sub:enumeration}). This is not always the case with other tools, such as GHC, which may only report a subset of the errors present in the code or stop at the first error they encounter. Goanna, however, always thoroughly identifies all type errors in the codebase. In the example of Fig.~\ref{fig:goanna-example-1} and Fig.~\ref{fig:goanna-example-2}, Goanna discovered the two errors included in the file. Clicking on the error selector at the bottom left will change the content of the debugging panel and the highlights of the text editor to reflect the cause of a different error (Fig.~\ref{fig:multi-error}). 

    \subsection{Type error grouping}  \label{sub:group}
    In addition to reporting multiple errors, Goanna also groups together type errors that might be treated as multiple separate errors by other tools. Goanna uses a novel approach (Section \ref{sub:grouping}) to ensure that type errors that are intuitively connected are grouped together. This means that Goanna does not overwhelm the programmer with an excessive number of redundant type errors. Instead, the programmer is presented with a concise list of errors that all can be assessed separately.

    \begin{figure}[htb!]
        \centering
        \includegraphics[width=\linewidth]{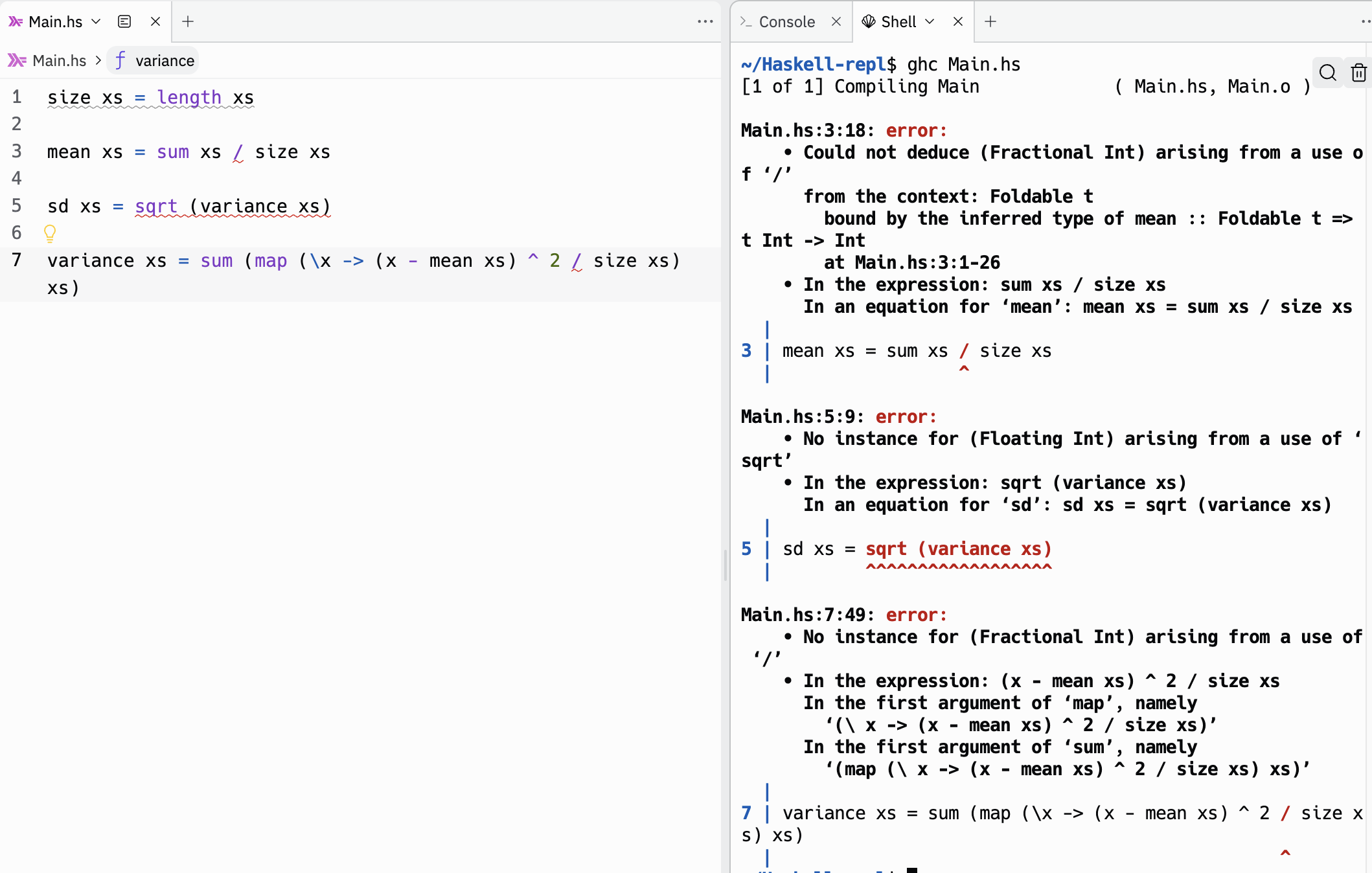}
        \caption{\textbf{Inspecting a defective Haskell Program (left) in relation to the error messages output by the standard GHC compiler (right)} -- 3 separate type errors are reported.  The editor (VS Code is used here) underlines the error locations reported in the messages, but all other contextual information must be understood from the error text.}
        \label{fig:grouping-ghc}
    \end{figure}
    
        \begin{figure}[htb!]
        \centering
        \includegraphics[width=\linewidth]{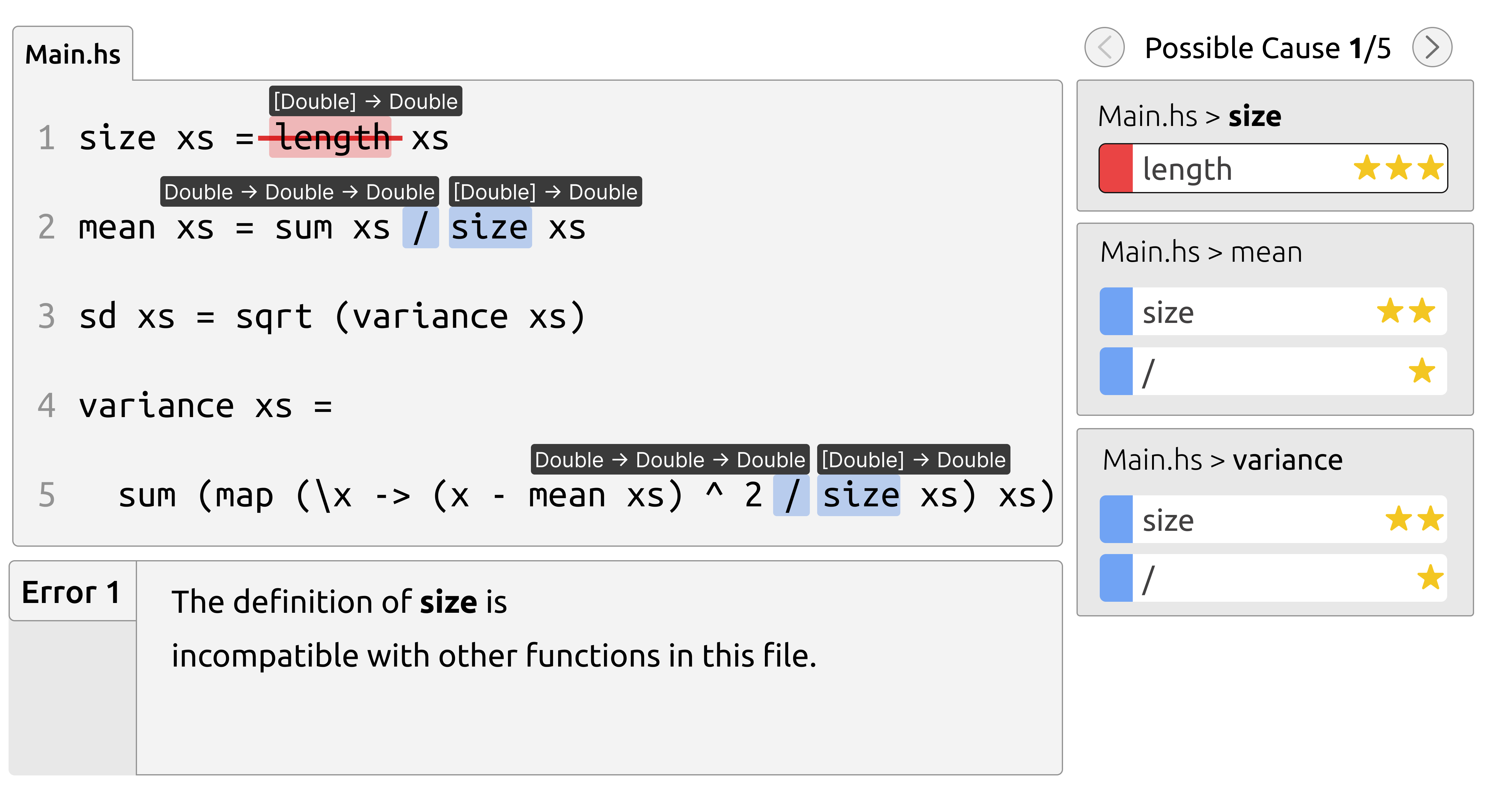}
        \caption{\textbf{Goanna's Error Grouping.} This error, although its potential offending parts appear in many declarations, is possible to fix in one place, i.e., by changing the definition of the \texttt{size} function on line 1. Therefore, Goanna reports it as a single error.}
        \label{fig:grouping-goanna}
    \end{figure}

    For instance, in Fig.~\ref{fig:grouping-ghc}, the functions \texttt{variance} and \texttt{mean} expect the final type of \texttt{size} to be a fractional value. However, the definition of \texttt{size} results in an integral value, which creates a conflict. While GHC shows three separate type errors, Goanna groups these interconnected errors into a single entity, as shown in Fig.~\ref{fig:grouping-goanna}. These errors can be addressed collectively, thus improving the efficiency of the programmer.

    \subsection{Discovering Potential Causes} \label{sub:suggesting}
    When a type error arises, Goanna-IDE shows a list of possible causes in the debugging panel. Each possible cause consists of one or more locations in the code that require modification to rectify the type error. Clicking on a possible cause activates it. The locations are highlighted in the text editor, as well as in the inlay type hints which suggest the suitable type expected for that code slice. In the debugging panel, the activated cause is outlined with a red icon, while others are marked with a blue icon. 

    The causes identified by Goanna are comprehensive. Goanna will take into account potential causes in expressions, pattern matchings, type annotations, and type class constraints. Consequently, programmers will generally find the real cause by exploring Goanna's diagnosis. Unlike most Hindley-Milner~\cite{Damas1982-sc} based type inferences, Goanna does not show a bias towards the unification order, thus avoiding the left-to-right bias \cite{Chen2014-ev}. 
    
    Note that Goanna's fixes are sufficient to resolve the type error. Traditional tools often reveal a set of partial locations of a type error, leaving programmers to realize later that additional adjustments are needed for a complete resolution. Goanna, however, offers fixes that encompass a complete set of changes necessary for a resolution.

    \subsection{Assessing Likelihood of Causes} \label{sub:conciseness}
    One challenge of Goanna's ``find all causes" approach is the number of ways an error can occur, which can sometimes become too large to be useful in practice. Goanna uses multiple techniques to intelligently sieve the list. For the remaining list, Goanna employs a few heuristics to rank their likelihood and inform programmers of which causes they consider first. 
    Goanna-IDE uses a star-based rating system to signal the ``likelihood'' of each cause. Three stars indicate the most likely cause, two stars and one star follows. 

    \subsection{Type Hints}\label{subset:type-hints}
    In addition to suggesting which part causes that type error, Goanna-IDE explains why this is inferred by using inlay type hints on necessary terms. The type hints are displayed as inlay decorations on top of respective fragments of source code. These type hints provide enough information for programmers to understand the type inference, and Goanna will leave out the terms that are irrelevant to the type error. Goanna's type hints are also dynamic to the selected cause. Programmers can observe how the inferred type of each term changes when the selected cause is changed. Many modern programming tools use inlay-type hints to support understanding, such as the Haskell Language Server~\cite{noauthor_2023-ot}. Most often, these tools will display type hints for all or none of the declarations. Unlike Goanna, these tools do not provide alternative sets of type hints for programmers to compare.
    
    \subsection{Cross-module type error debugging}\label{subsec:cross-module-type-error-debugging}
\begin{figure}[htb!]
        \centering
        \includegraphics[width=\linewidth]{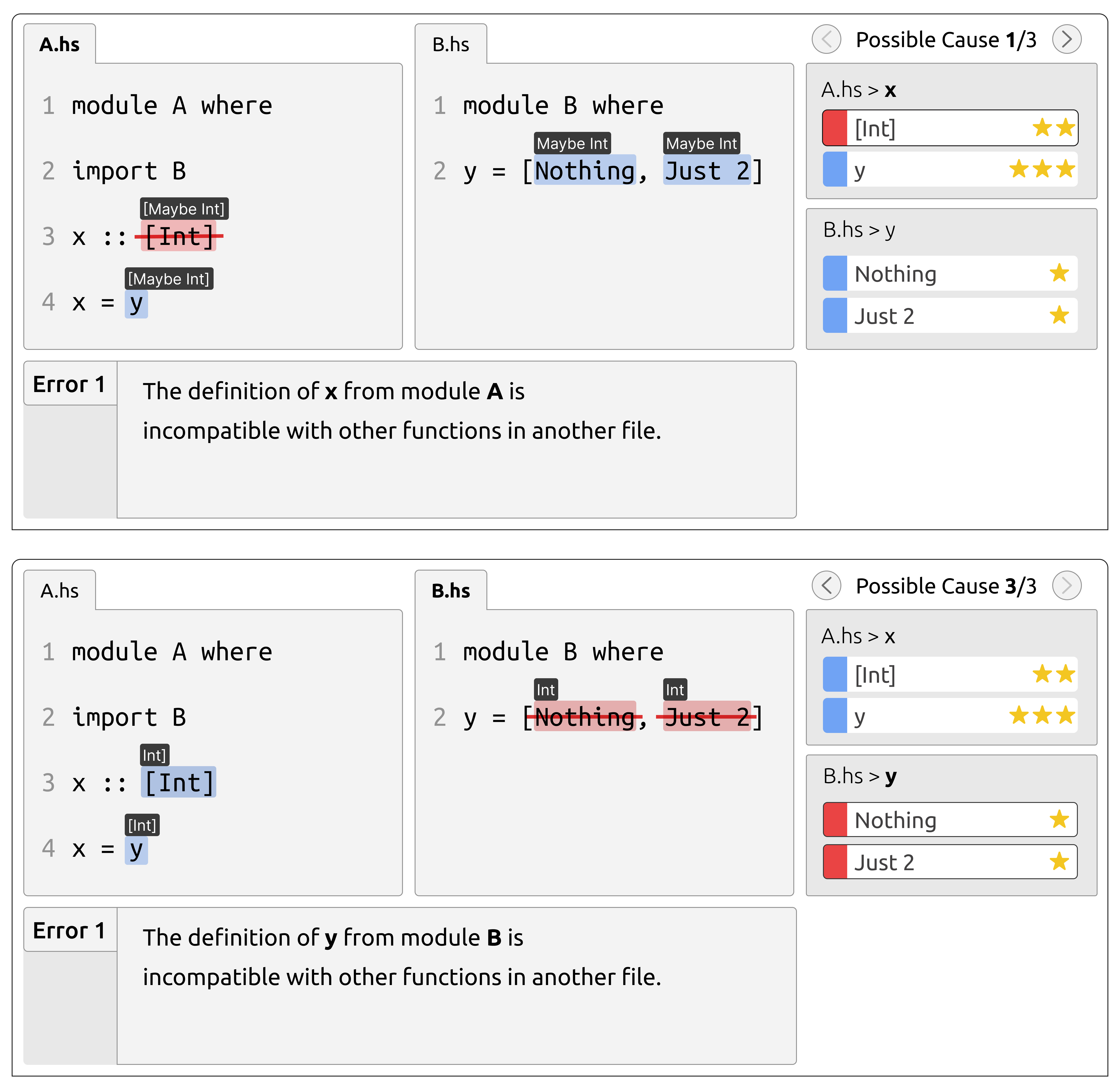}
        \caption{\textbf{Debugging a cross-module error in Goanna.} In this error, potential defects may appear in either module \texttt{A} and \texttt{B}. Goanna suggests 3 potential causes and fixes: 1) Change the type annotation of \texttt{x} to \texttt{[Maybe Int]} (Top). 2) Change the y variable on line 4 of module \texttt{A} to an instance of \texttt{Int}. 3) Change both the elements in the list literal in module \texttt{B} (Bottom), hence affecting the type of \texttt{y}. Clicking on each potential cause in the debugging panel results in different highlights and type hints in the editor panel.}
        \label{fig:goanna-cross-module}
    \end{figure}

    When encountering a type error that spans multiple modules, Goanna-IDE will group the potential causes indicated by their module and declaration block. Clicking on any possible cause location will focus the editor on the corresponding module (Fig.~\ref{fig:goanna-cross-module}). Goanna is the first tool to introduce cross-module type debugging. The way Goanna presents cross-module type errors is analogous to how run-time errors are presented in most programming languages. When encountering a run-time error, most programming environments show a call stack containing multiple file paths, and programmers can choose which file to start investigating. Often, programmers choose to start from the file authored by themselves instead of library files. Goanna uses this mental model to group potential locations that cause a type error by module and definition blocks. 
    
    \section{Goanna Implementation} \label{sec:implementation}
    Goanna comprises 3 phases: constraint generation, MCS enumeration, and post-analysis. In the constraint generation phase, Goanna walks the abstract syntax tree and collects constraints. In the MCS enumeration phase, Goanna enumerates through all MCSes. Lastly, in the post-analysis phase, Goanna applies multiple optimization techniques to reduce the number of MCSes, group MCSes by common properties, and sort them according to a selection of heuristics.

    \subsection{Haskell Coverage}
    Goanna supports a wide and growing range of Haskell 2010 language syntax \cite{Simon_Marlow2010-lg}. At the time of writing, fully supported features include module import/export, qualified imports, import hiding, do notation, algebraic data type, new type, type synonym, type class, operator sectioning, range expression, N+K pattern, with record syntax, and list comprehension are under development. Goanna does not yet support type features enabled through language extensions. However, they are also on the roadmap. A detailed and updated feature coverage list can be found in \cite{Anonymous2023-rp}.

    \subsection{Constraint Generation} \label{sub:translation}
    Goanna uses the abstract syntax tree of the original Haskell program and translates it into a constraint program by modeling how types are defined and used. Goanna does not restrict which constraint language and solver should be used. The only requirement is that Goanna can assert whether a subset of the constraints is still feasible by calling a provided \texttt{solve} function during the MCS enumeration phase. In our implementation, we generate portable Prolog predicates \cite{Wielemaker2011-sr}. The \texttt{solve} function executes a predefined predicate \texttt{type\_check/0} that tests all generated predicates. We used the standard Prolog notation \texttt{name/arity} here when referring to Prolog predicates, as a Prolog predicate is identified by the combination of both attributes.

\begin{figure}[htb!]
        \centering
        \includegraphics[width=\linewidth]{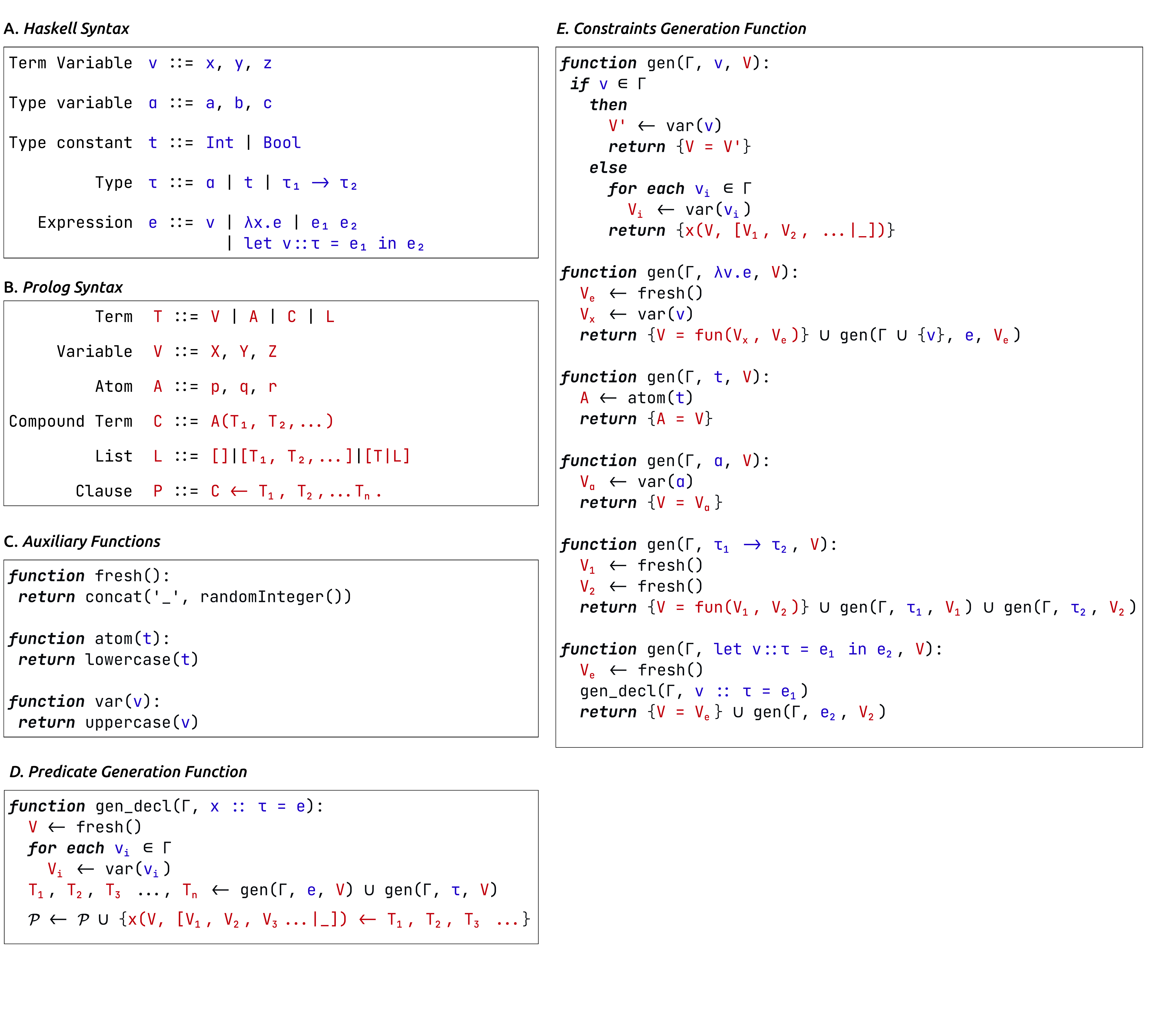}
        \caption{Goanna's Constraint Translation Rules (Simplified)} 
        \label{fig:translation}
    \end{figure}

    For a simplified Haskell syntax shown in Fig.~\ref{fig:translation}.A, we generate a list of Prolog predicates in the language shown in Fig.~\ref{fig:translation}.B. We use three auxiliary functions during the constraint translation process (Fig.~\ref{fig:translation}.C) to generate Prolog variables for future unification. \texttt{fresh} makes a unique unbound Prolog variable. \texttt{var} takes a Haskell identifier name and returns a Prolog variable. Naively, this can be done by turning it to uppercase. \texttt{atom} takes a Haskell type constant/constructor name and returns a Prolog atom. Naively, this can be achieved by turning it into lowercase. We keep track of the local variable names in a list $\Gamma$ containing the Haskell variable names. A global variable $\mathcal{P}$ is defined to store the list of predicates that are being generated. To clarify, all \textcolor{blue}{source Haskell syntax} are in blue, and all \textcolor{red}{generated Prolog syntax} are in red. 
    
    Two sets of generation rules apply to different syntax nodes and output Prolog source code. Predicate generation rules (Fig.~\ref{fig:translation}.D) take Haskell declarations as input and output Prolog predicates. For example, a top-level Haskell declaration \texttt{f = 2}  may generate a predicate \texttt{f(V, \_)} 	$\leftarrow$ \texttt{V = int}.
    
    Constraints generation rules (Fig.~\ref{fig:translation}.E) take a Haskell expression node or type node and a Prolog variable \texttt{V} as input and output a list of Prolog terms. These terms attempt to unify the inferred type of such node with the provided Prolog variable \texttt{V}.
    
    \begin{figure}[htb!]
        \centering
        \includegraphics[width=0.6\linewidth]{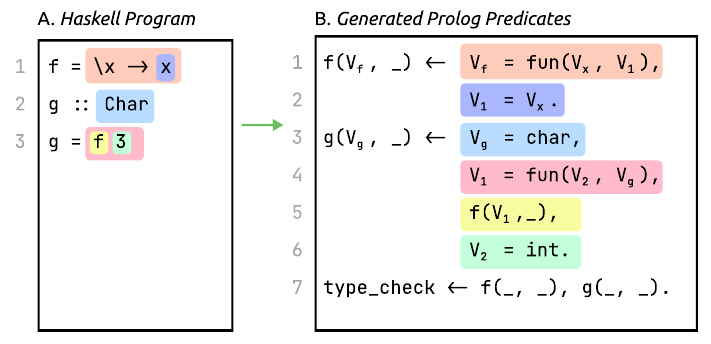}
        \caption{\textbf{An example of Goanna constraint generation.} For the Haskell functions \texttt{f} and \texttt{g}, Goanna generates the predicates \texttt{f/2} and \texttt{g/2}. Each subgoal of \texttt{f/2} and \texttt{g/2} is generated from a corresponding part of the Haskell program. In a predefined predicate \texttt{type\_check/0}, the subgoals \texttt{f(\_,\_)} and \texttt{g(\_,\_)} are added. Running the goal \texttt{type\_check} will return whether the program is well-typed. In this particular example, this will return \texttt{false}. We used standard Prolog notation \texttt{name/arity} here when referring to Prolog predicates, as a Prolog predicate is identified by the combination of both attributes. 
}
        \label{fig:translation-example}
    \end{figure}

    An example of such translation can be found in Fig.~\ref{fig:translation-example}. In the Haskell program (Fig.~\ref{fig:translation-example}.A), two functions are declared: \texttt{f} and \texttt{g}. This will generate two corresponding Prolog predicates \texttt{f/2} and \texttt{g/2}. In the actual implementation of Goanna, the generated predicates would be \texttt{f/6} and \texttt{g/6}. The extra arguments are added to perform a series of logistic tasks for realistic type checking, such as breaking recursive calls and collecting type-class constraints. In a predefined predicate \texttt{type\_check/0}, the subgoals \texttt{f(\_,\_)} and \texttt{g(\_,\_)} are added. Executing the top-level goal \texttt{type\_check} in a Prolog environment will get a result of \texttt{false}, signaling that the source program is ill-typed.

    \subsection{MCS enumeration} \label{sub:enumeration}
    After the constraint generation phase, Goanna obtains a list of constraints derived from the source code and can query the feasibility of any subset of the constraint system by calling the \texttt{solve} function. Using a known algorithm \cite{Liffiton2016-xi}, Goanna then derives some useful subsets of the constraint system through the MCS enumeration. We refer to the complete set of constraints as a constraint system $C$. When we use the word subset without specifying the corresponding superset, it can be inferred to be the subset of the constraint system $C$. We list these subsets obtained from the MCS enumeration and give their type-theoretic interpretation.

    – A minimal unsatisfiable subset (MUS) $M$ of a constraint system $C$ is a subset $M \subseteq C$ such that $M$ is unsatisfiable and $ \forall{c} \in M : M \setminus \{c\}$ is satisfiable. An MUS can be seen as a minimal explanation of the infeasibility of the constraint system. MUSes have been used extensively, mostly in combination with programming slicing, as a means to explain type errors. An MUS of type system constraints encodes a path of reasoning connecting all evidence from one location of the conflict to another. Goanna uses the set of all MUSes to group related type errors.

    – A minimal correction set (MCS) $M$ of a constraint system $C$ is a subset $M \subseteq C$ such that $C \setminus M$ is satisfiable and $\forall{S} \subset M : C \setminus S$ is unsatisfiable. MCSes are so named because their removal from $C$ can be seen to “correct” the infeasibility. In an ill-typed program, an MCS can be seen as the ``cause" of a type error; removing any MCS will result in the system being well-typed. Goanna uses an MCS to represent potential causes of a type error. Each MCS contains the set of locations that need to be changed to fully resolve the type error.
    
  – A maximal satisfiable subset (MSS) $M$ of a constraint system $C$ is a subset $M \subseteq C$ such that M is satisfiable and $\forall{c}\ in\ C \setminus M:M\cup\{c\}$ is unsatisfiable. The definition of an MSS is symmetric to that of a MUS, with `satisfiable' and `unsatisfiable' swapped along with maximal for minimal. MCS and MSS are complement sets of each other. In an ill-typed program, an MSS can be seen as the resulting typing environment if a type error is fixed by excluding the MCS. Goanna uses an MSS to provide type hints for the program even when it is ill-typed.
  

     \begin{figure}[htb!]
        \centering
        \includegraphics[width=\linewidth]{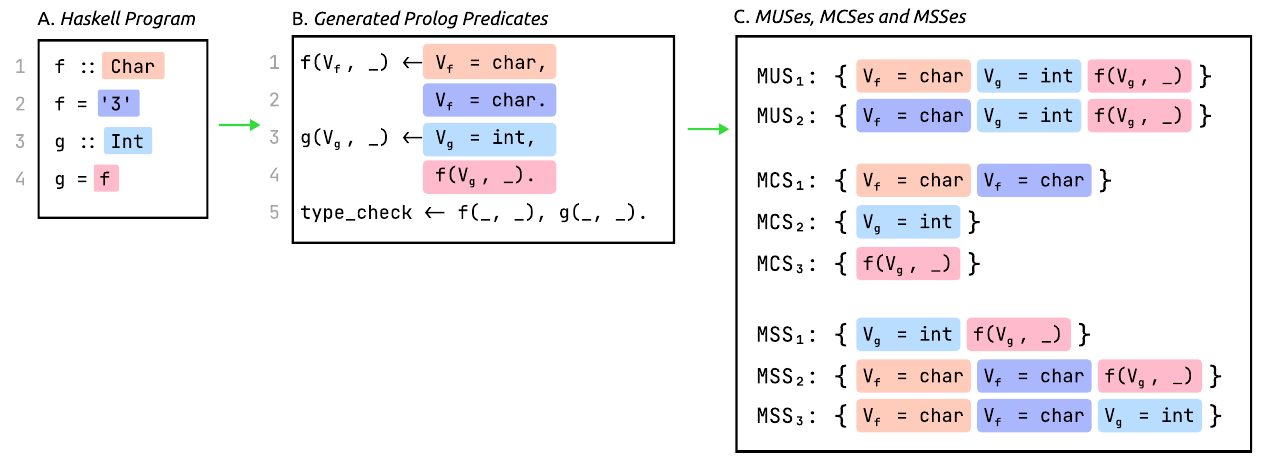}
        \caption{\textbf{An example of Goanna MCS Enumeration.} From the set of constraints (B) generated from the Haskell program(A), Goanna obtained 2 MUSes, 3 MCSes, and 3 MSSes. }
        \label{fig:enumeration-example}
    \end{figure}
    
   For the example in Fig.~\ref{fig:enumeration-example}, Goanna's MCS enumeration system identifies 2 MUSes, 3 MCSes, and 3 MSSes. Following the 3 MCSes, Goanna reports 3 potential causes of the type error: the type annotation and function definition in \texttt{f} (from $MCS_1$), the type annotation alone in \texttt{g} (from $MCS_2$), and the function definition alone in \texttt{g} (from $MCS_3$).

	\subsection{Post-Analysis}
	Three types of post-analyses are used to improve the quality of error diagnoses: type error grouping, cause ranking, and cause reduction. 
    \subsubsection{Type Error Grouping} \label{sub:grouping}

    Type error grouping is a novel feature provided by Goanna. Conventionally, in a type error slicing approach, a type error is represented by a minimal unsatisfiable subset (MUS). With multiple MUSes available, we have the knowledge to be more precise about an ill-typed program. We propose a novel method for representing type errors that aligns more closely with their colloquial meaning.

	Let $U$ denote the set of all Minimal Unsatisfiable Subsets (MUSes) and $C$ the set of all Minimal Correction Sets (MCSes). We define an undirected graph $G$, where each vertex in $G$ corresponds to a minimal unsatisfiable subset $u_i \in U$, and the edges of $G$ connect pairs of MUSes $u_i$ and $u_j \in U$ if their intersection is nonempty. The set of all connected components $D$ in $G$ represents the set of all type errors. For each $d_i \in D$, let $l_i = \bigcup v_i$, where $v_i$ is the set of vertices in $d_i$. $l_i$ is the set of all constraints local to this type error. Define $C_i = \{ x \mid \forall c \in C, x = c \cap l_i \}$ as the set of all MCSes that are local to this type error.

\begin{figure}[htb!]
        \centering
        \includegraphics[width=0.6\linewidth]{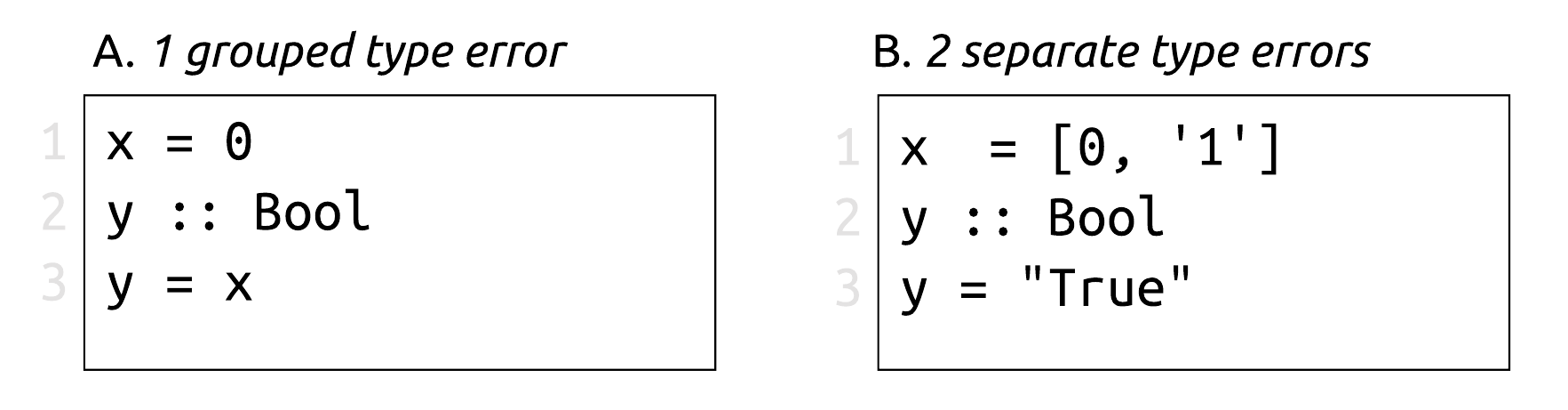}
        \caption{\textbf{Goanna's type error grouping.} The ill-typed program on the left contains a single type error, because it can be fixed by a minimal set of syntax changes. For example, fixing it by changing the literal \texttt{0} on line 1 to \texttt{True} or \texttt{False}. This edit contains a single location, so there exists no smaller edit that can fix \texttt{x} or \texttt{y} alone. The program on the right contains two type errors because \texttt{x} or \texttt{y} can be fixed separately. For example changing \texttt{0} to \texttt{'0'} on line 1 fixes \texttt{x} alone. }
        \label{fig:grouping-example}
    \end{figure}

    This can be intuitively thought of as follows: two type errors can be grouped together if they cannot be fixed independently through modifying a minimal set of locations for each. For instance, Fig.~\ref{fig:grouping-example}.A shows one connected type error, where there are two fixes available: change \texttt{0} on line 1 to a Boolean type, change the type annotation on line 3 to \texttt{Integer}, or change the assignment of \texttt{y} to a different expression. Choosing either one will result in both \texttt{x} and \texttt{y} being inferred to have a valid type.

    However, in Fig.~\ref{fig:grouping-example}.B, we can fix \texttt{x} or \texttt{y} separately. For example changing \texttt{0} to \texttt{'0'} on line 1 fixes \texttt{x} alone. Conversely, changing the type annotation on line 2 to \texttt{String} will fix \texttt{y} and leave x still ill-typed. In this case, there are two separate type errors that should not be grouped.

	In practice, the grouping of type errors provides a sense of the ``effective area" of a type error. Programmers are commonly bewildered by the fact that changing one place of the program causes an error in a seemingly unrelated area. This property allows Goanna to be helpful in refactoring a known correct program. When a programmer change the definition of one funtion, Goanna will show all the locations that require further changes in order for the program to be successfully refactored. This works because all the further changes belong to the same type error group, because a single syntax change -- reverting the initial change -- will result in the program being well-typed once more.
	
	More specifically, to Goanna, type error grouping provides an effective means to reduce the number of causes. In an ill-typed program with $m$ errors, each having $n$ potential causes, will result in $nm$ total causes. Dividing these causes into separate errors that align correctly with intuition is the most important technique to enhance Goanna's error reporting.

    \subsubsection{Cause Ranking} \label{sub:ranking}
     In Goanna-IDE, when a list of potential causes is displayed, Goanna-IDE also shows the top 3 ``likely" causes according to the ranking heuristics. This is very helpful because programmers will have a starting point for the investigation. We have identified several efficient heuristics for ranking suggestions. Although no single heuristic ensures universal applicability,  a healthy combination of all the listed heuristics delivers satisfactory results across a broad spectrum of error scenarios.

    \paragraph{Number of Error Locations.}
    Causes comprising fewer locations are prioritized and presented earlier in the list. For example: \begin{figure}[htb!]
        \centering
        \includegraphics[width=0.4\linewidth]{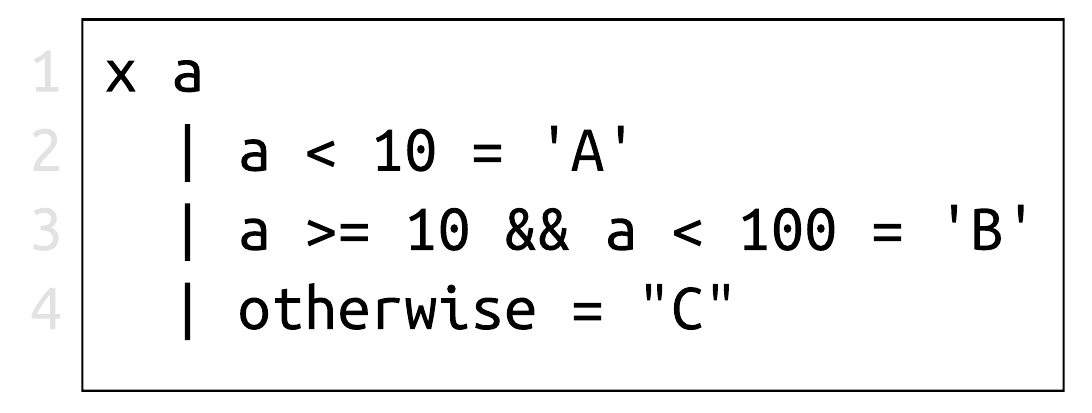}
        \caption{\textbf{Goanna prefers causes with fewer error locations.} In this ill-typed program, Goanna chooses to give the cause \texttt{"C"} on line 4 a higher likelihood because it contains a single location. The other cause contains 2. }
        \label{fig:loc-count}
    \end{figure}

    In the example in Fig.~\ref{fig:loc-count}, there exist two possible fixes: 1) changing \texttt{'A'} and \texttt{'B'} to the string type, and 2) changing \texttt{"C"} to the \texttt{Char} type. As the latter fix affects only one location (as opposed to 2 in the former), it is assigned a higher ranking and appears earlier in the list.

    \paragraph{Change specificity.}
	Another useful heuristic is to encourage the cause whose fix will result in every surrounding term to be as concrete as possible.

   \begin{figure}[htb!]
        \centering
        \includegraphics[width=0.4\linewidth]{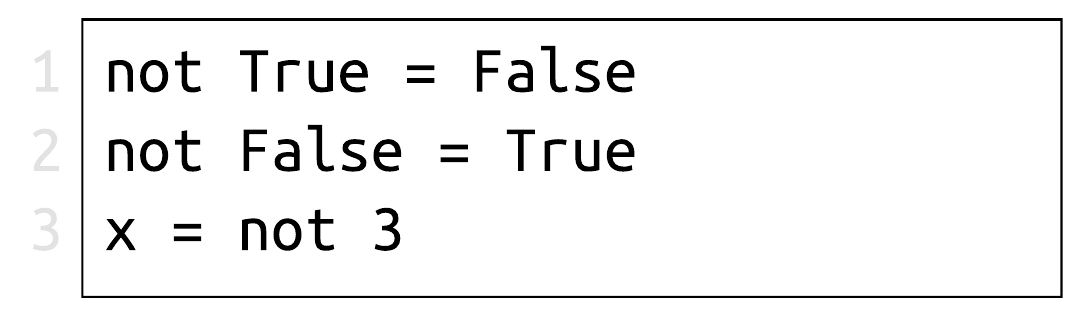}
        \caption{\textbf{Goanna prioritize causes whose resolutions lead to more concrete type assignments.} In this example, change \texttt{3} will result in \texttt{x} to have type \texttt{Bool}. Alternatively, \texttt{x}'s type will be unknown after changing \texttt{not} on line 3. Goanna prefers the former.} 
        \label{fig:specificity}
    \end{figure}

    In the example in Fig.~\ref{fig:specificity}, Goanna can suggest two potential causes and fixes. First, change the integer literal \texttt{3} to a \texttt{Bool} type. Second, change the function \texttt{not} to a different function that accepts an integer as input. The second fix results in the variable \texttt{x} having a less concrete type. Indeed, \texttt{x} can have any type if we do not limit what function to replace \texttt{not} with. Goanna prioritizes the first cause over the second. 

    \paragraph{Error span.}
    Goanna prioritizes the potential causes whose corresponding locations are clustered within fewer function definitions and lowers the likelihood of those whose corresponding locations are spread across multiple definitions or even multiple files.

    \subsubsection{Cause reduction} \label{sub:optimization}
    
    We employ three techniques to elevate the clarity of the suggestion list: reduction of constraint count, elimination of over-fitting resolutions, and elimination of redundant fixes.

    \paragraph{Minimize Constraint Count.}
    The number of constraints directly influences the time complexity associated with enumerating the Minimum Correction Subset (MCS). By merging multiple constraints into a singular one, we can reduce the total count of constraints and, in turn, improve the performance of the enumeration. Goanna perform the merging of consitraints during the generation phase. Yet, this approach requires careful application, as it could potentially lead to unsolvable situations or propose infeasible fixes, as the combined locations in the source code must either all contribute to the type error or none do. An effective application of this technique is to merge all constraints created by sub-expressions in a type signature.

   \begin{figure}[htb!]
        \centering
        \includegraphics[width=\linewidth]{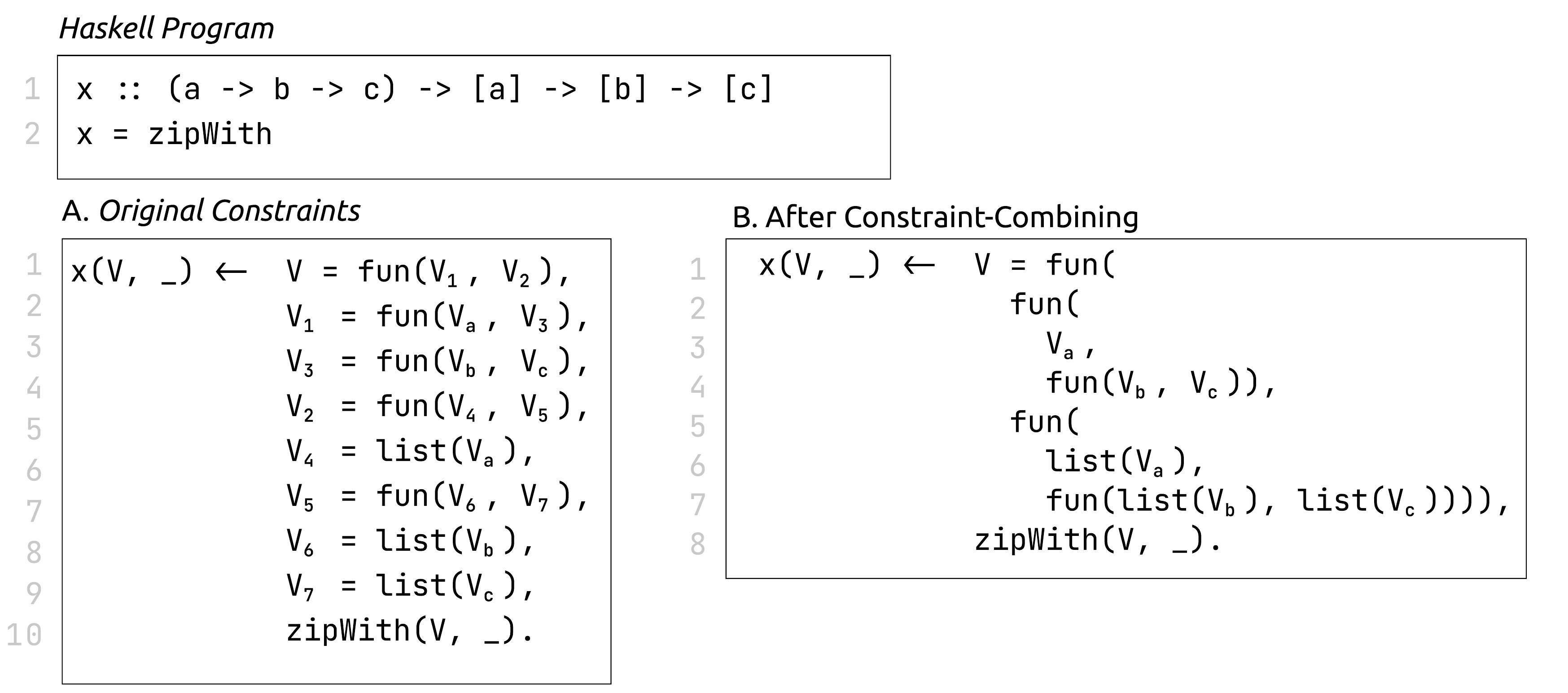}
        \caption{\textbf{Combine constraints in type signature.} For a simple Haskell program (top),  without any optimization, Goanna generates 10 constraints (bottom left), indicated by 10 subgoals in the predicate. By combining the constraints in the type signature, Goanna produces 2 constraints (bottom right).}
        \label{fig:combine-constraints}
    \end{figure}

    Consider the example in Fig.~\ref{fig:combine-constraints}. Without optimization, this code would spawn 10 constraints. However, applying this optimization can achieve an equivalent outcome with merely 2 constraints. Notably, this optimization forfeits the capability to suggest fixes for sub-expressions within a type signature, a decision that warrants some consideration but, in our experience, pays off.

    \paragraph{Elimination of Over-Fitting Resolutions.}
    In general, every syntax node in the source code generates one or more constraints. This includes structural nodes such as function applications. However, very often, suggesting that the user should modify the entire function application expression is not particularly instructive when changing one of the arguments fixes the type error as well. Disabling suggestions for overfitting solutions improves the clarity of the suggestion list and enhances the speed of MCS enumeration.

    \paragraph{Elimination of Redundant Causes.}
    Goanna iterates over the possible causes and removes the ones that fail to provide new insights. If all locations in a cause suggestion have already been covered in preceding suggestions, subsequent suggestions that merely rearrange these locations in different combinations can be omitted.
    
   \begin{figure}[htb!]
        \centering
        \includegraphics[width=0.9\linewidth]{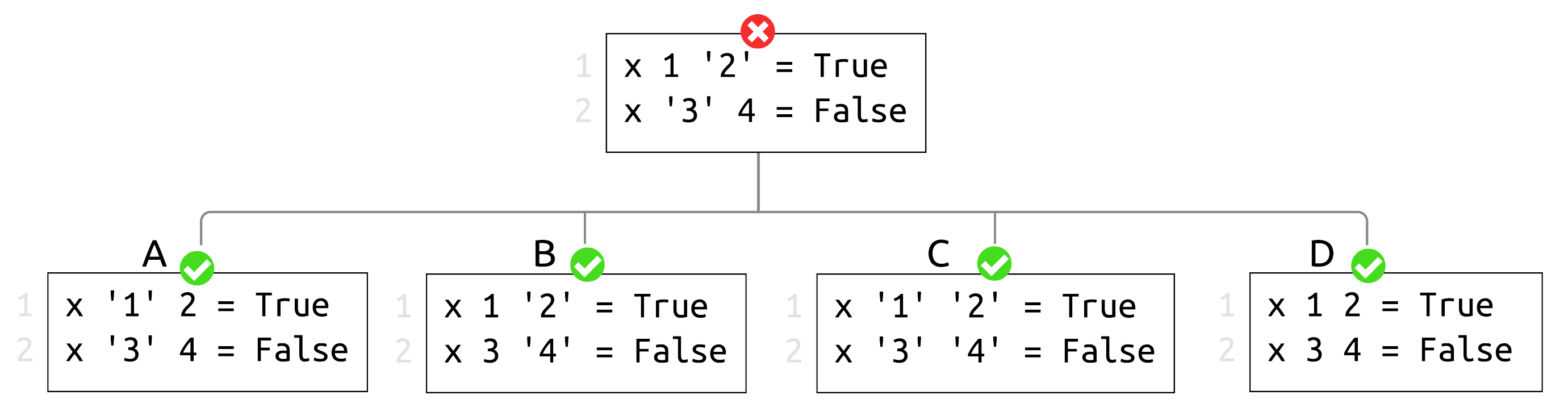}
        \caption{\textbf{The number of potential causes can grow exponentially.} In the ill-typed Haskell program (Top), there are 4 different ways (Bottom) to fix the type error. It is not hard to see this growth is exponential, and showing all the alternatives is not helpful.}
        \label{fig:reduction-example}
    \end{figure}
    
    Consider the example in Fig.~\ref{fig:reduction-example}. Without knowing the programmer's true intention, Goanna can provide four ways to fix the issue shown at the bottom. However, closer inspection reveals that after the first two suggestions, we no longer unearth new insights. Therefore, they can be removed to enhance the clarity of the suggestion list. 
    Note that this can remove the correct answer (say D), but if the programmer uses part of the (A) to make the fix, the revised type error will include the correct fix. 
    
    Removing superfluous MCS-based suggestions that recycle different permutations of the same set of locations is an instantiation of the Set Cover Problem (SCP). The problem can be rephrased as finding the minimal number of MCSes that cover all the potential locations that could cause the type error. Many approaches solve the SCP \cite{Caprara2000-vw}, including eager algorithms, linear programming, and heuristic-based algorithms. Generally, we found that all of these approaches find the minimal cover of type errors efficiently. Goanna uses the OR-tools \cite{Google_Developers_undated-ew} for this task.
       
    \section{Evaluation} \label{sec:evaluation}

     We want to answer the following key research questions about our Goanna prototype:

    \begin{itemize}
        \item RQ1. Does Goanna provide a more accurate type error diagnosis compared to traditional tools?
        \item RQ2. Does Goanna provide a concise list of suggestions?
        \item RQ3. How efficiently does Goanna compute error diagnoses?
    \end{itemize}

    \subsection{Experiment Design} \label{sub:dataset}

    To evaluate Goanna, we extracted a collection of defective Haskell programs (N=86) from Haskell online discourse since 2018, each containing one or more type errors. The communities we searched include StackOverflow (32), Haskell on Reddit (20), and Haskell Discord Channel (34), as these are the top discussion channels for Haskell users \cite{Fausak2022-zf}. During the search process, we looked for online discussions where the authors encountered type errors in their Haskell programs and asked for help. We selected only the questions that had been answered, and furthermore, the answers had been accepted by the original author. We extracted the defective Haskell programs and the accepted answers as the oracle solution. The length of these programs ranges from 1 line to 64 lines of code (mean=20, median=20). These programs span a variety of subjects, including basic syntax (14 files), lists (28 files), tuples (5 files), algebraic data types (22 files), higher-order functions (17 files), monadic operations and do notation (9 files), type classes (6 files), and built-in/library functions (24 files). The distribution of themes generally aligns with the breakdown of the different causes of the type errors from Tirronen's study \cite{Tirronen2015-nr}.
    
    	For each metric in the evaluation, Goanna is compared with Glasgow Haskell Compiler (GHC) \cite{Ben_Gamari2022-bs} and Helium \cite{Helium4Haskell2023-kk}. GHC was chosen as the baseline because of its established reputation in the Haskell community, its wide capability, and its great efficiency in working with Haskell projects. Helium is acknowledged to produce high-quality error messages \cite{Heeren2003-kd}. The experiments were run with GHC 9.4 and the standalone Helium compiler version 1.8.

 	\subsection{RQ1. Accuracy}\label{sub:eval-accuracy}
\begin{figure}[htb!]
        \centering
        \includegraphics[width=\linewidth]{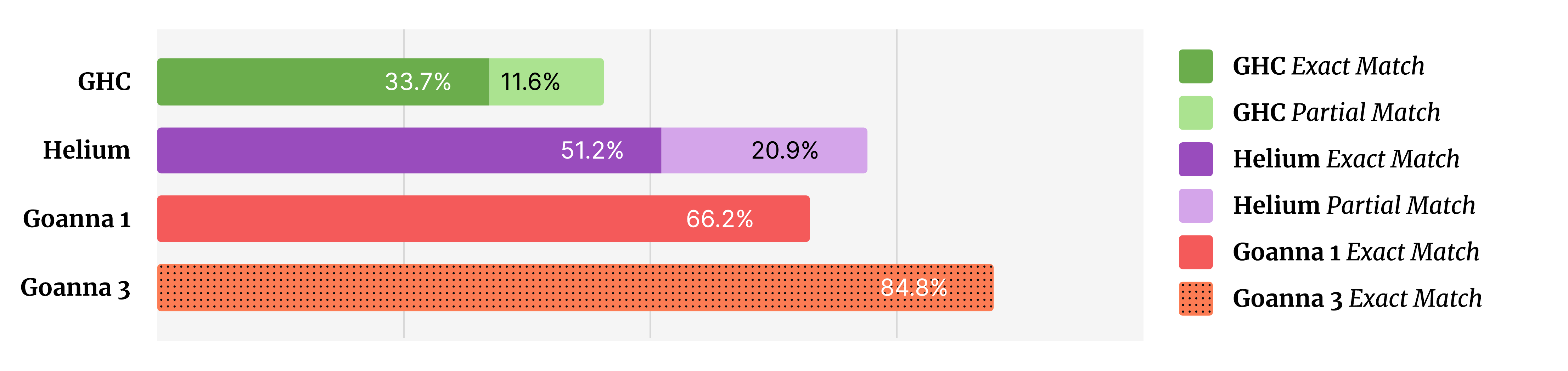}
        \caption{\textbf{The percentage of diagnoses that match the accepted answers.} }
        \label{fig:accuracy}
    \end{figure}

For each program in our dataset, we compared the error diagnosis of each tool with the accepted answer. We consider the diagnosis accurate only if its suggested fix matches the accepted answer. We consider the diagnosis partially accurate if the tool's diagnosis is part of a larger set of locations that make up the intended cause or if the diagnosis addresses one of the multiple errors. Because Goanna provides a comprehensive list of possible causes, it is very likely that all sensible fixes are included. In this evaluation, we only consider the top-1 suggestion (Goanna 1) and top-3 suggestions (Goanna 3).  From the graph in Fig.~\ref{fig:accuracy}, GHC's accuracy is the least performant among all tools. Goanna 1's, although lower than Helium (72.1\%) in partial accuracy, is higher than Helium (51.2\%) when considering only diagnoses that fully match the accepted answer. \textbf{Goanna 3 has the best accuracy (84.8\%)}, higher than the partial accuracy of other tools.
  
 	\subsection{RQ2. Conciseness}\label{sub:eval-conciseness}
\begin{figure}[htb!]
        \centering
        \includegraphics[width=\linewidth]{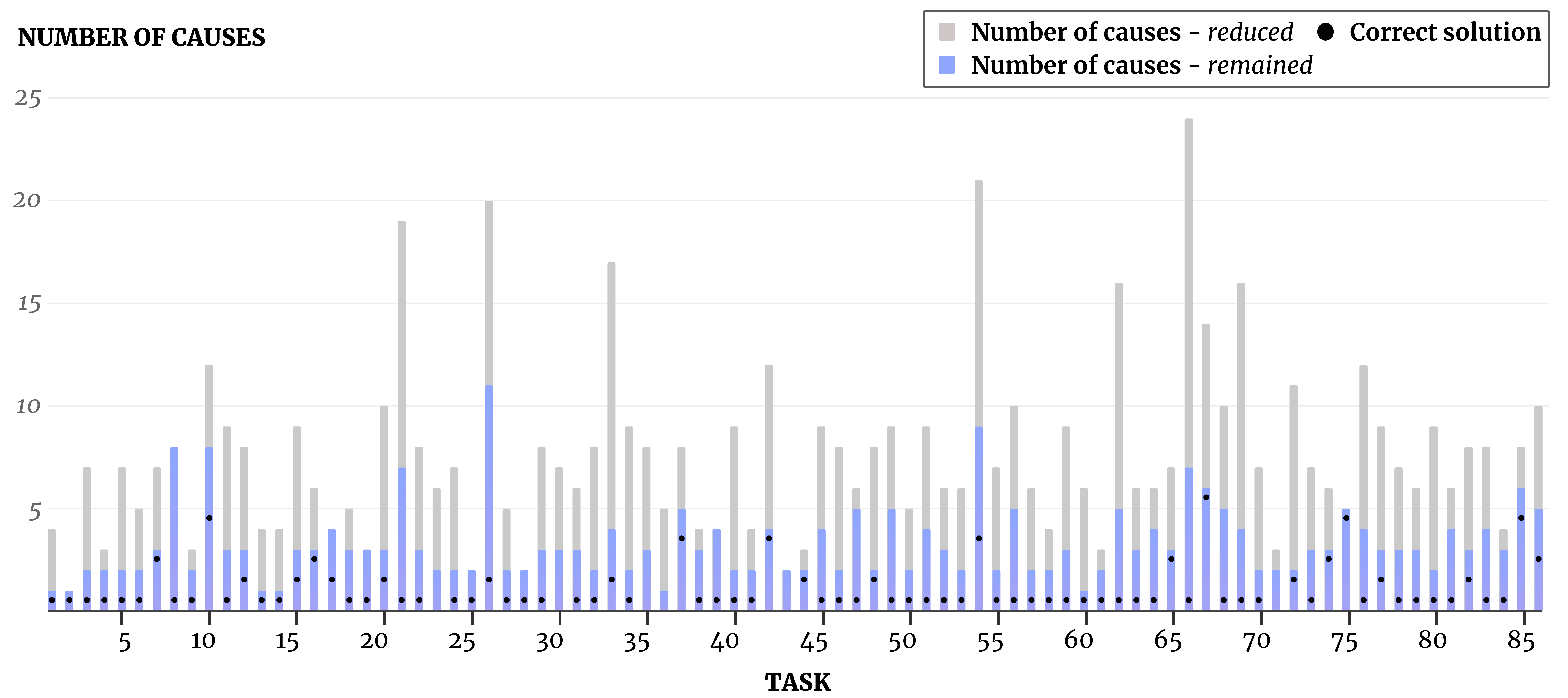}
        \caption{\textbf{The number of potential causes identified by Goanna.} }
        \label{fig:conciseness}
    \end{figure}

 Using Goanna requires users to cherry-pick from a list of possible causes. It will severely reduce usability if the list is too long. To evaluate Goanna's conciseness, we counted the number of suggestions provided by Goanna in all the tasks.  We also indicate where the accepted answer is. Additionally, we included a baseline of Goanna with all cause-reduction features disabled. As shown in Fig.~\ref{fig:conciseness}, \textbf{Goanna manages to effectively condense its suggestion list}, on average providing a short list of suggestions (mean=3.29, median=3.0) for each type error. Furthermore, on average, the accurate cause can be found within the top 2 suggestions (mean=1.63, median=1.0) to find the correct cause and fix. It also shows in Fig.~\ref{fig:conciseness} that Goanna's cause reduction strategies are effective; on average, 51\% of total causes are reduced to gain clarity. One unusual observation is that in 3 tasks, Goanna failed to include the correct solution. The current version of Goanna is ineffective in making the correct suggestions for these type errors. We discuss this in section \ref{sec:edge-case}. 
 
    \subsection{RQ3. Performance} \label{sub:eval-performacne}

\begin{figure}[htb!]
        \centering
        \includegraphics[width=\linewidth]{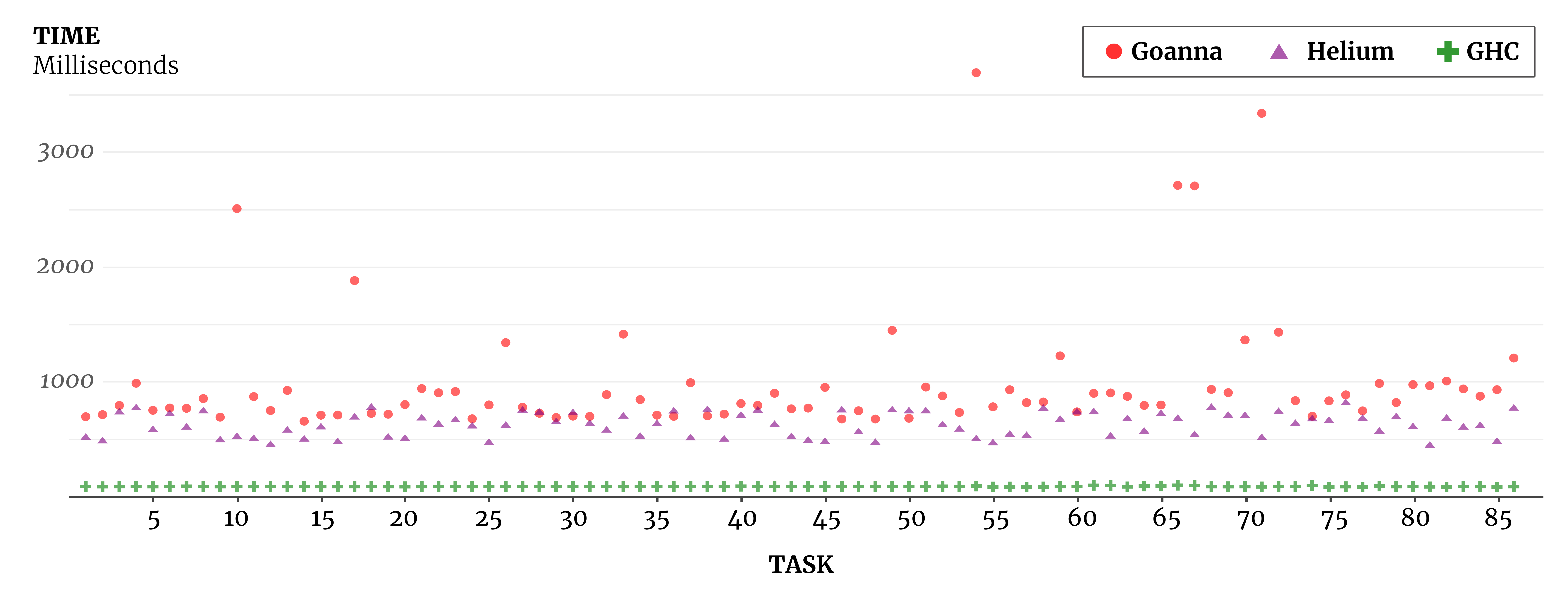}
        \caption{\textbf{The time it takes to type check and diagnose each program.} }
        \label{fig:performance}
    \end{figure}

    Goanna's performance largely depends on the MCS enumeration phase. Enumerating all MCS is computationally expensive. We experimentally compared the time it takes for Goanna to provide a complete error diagnosis for each task with GHC and Helium. From the data shown in Fig.~\ref{fig:performance}, we can see that \textbf{Goanna is slightly slower than Helium}~(Goanna: mean=0.98 seconds, median=0.83 seconds; Helium: mean=0.63 seconds, median=0.63 seconds). Goanna is approximately 10 times slower than GHC (mean=0.09 seconds, median=0.09 seconds) but greatly outperforms GHC (see Figure \ref{fig:accuracy}). One important pattern is that Goanna's response time varies more than that of other tools (Goanna SD = 0.55, GHC SD = 0.00, Helium SD = 0.10). It can be seen from Fig.~\ref{fig:conciseness} and Fig.~\ref{fig:performance} that the tasks that Goanna struggles most with are the ones that have significantly more potential causes. Multiple avenues exist to mitigate this delay, and we will discuss them in section \ref{sec:future-work}.
    
    \subsection{Threats to Validity}

    \paragraph{\textbf{Selection of the dataset}}
    Our dataset is limited in its number of cases and variety of type errors. This is due to the challenge of finding programs that contain type errors. Unlike runtime errors, which can be mined from code repositories and version control histories, type errors in Haskell can be detected by the compiler tool, and ill-typed programs are usually fixed before the changes are committed to the version control systems. In addition, we employed two selection methods. First, the error is indeed a type error. We test this by running the original program in GHC and checking if it indeed triggers a type error. Otherwise, the program is discarded, for instance, if it contains only parsing errors or runtime errors. Second, we discarded type questions where the main error relies on third-party libraries.

    \paragraph{\textbf{Measurement of performance}} Performance on Goanna and GHC was measured on a Linux virtual machine with a 3.1 GHz Processor and 2GB RAM. In practice, complex systems like this may perform differently depending on hardware and software configurations. Although we were unable to extract the performance profile of each tool on different platforms and operating systems, we chose hardware with abundant resources and up-to-date software dependencies. During our performance measurement, neither CPU usage nor memory usage was fully stressed. Additionally, GHC was run with the ``-fno-code'' flag enabled to limit its usage to type-check only rather than generating additional low-level code.

    \section{Discussion} \label{sec:discussion}


    \subsection{Strengths}
    Goanna demonstrates a notable improvement over existing Haskell type error detection and repair tools. Compared to traditional type-checking tools such as GHC, Goanna delivers improved error detection accuracy and flexibility to inspect different potential causes. The data suggest that users typically need to consider only 2-3 suggestions to achieve a satisfactory result.

    \paragraph{\textbf{Accurate suggestions}} Goanna is able to identify causes for Haskell type errors more accurately than GHC and Helium. We attribute this to a few factors. First, Goanna is the only tool capable of suggesting the type error in multiple nodes. In a study~\cite{Wu2017-eb} of over 2700 ill-typed Haskell programs, only 35\% of the type errors were caused by a single location. However, most of the type debugging tools only focus on single-location causes due to their technical limitation or to avoid high computational cost. Second, Goanna is the only tool that provides alternative causes of a type error. We could see that although Goanna-1 is not as accurate as Helium when accepting partial fixes, Goanna-3 surpasses Helium in accuracy. This translates into accurate type error identification at the cost of presenting the top 3 answers from Goanna instead of one.

    \paragraph{\textbf{Goanna provides contextual information}}
    Goanna provides type information for relevant terms to support each of its claimed causes. In traditional tools, the type-level information is often incomplete or completely discarded. In runtime error debugging, one of the most common features is inspecting the values of different expressions of the program. It would be ineffective if the run-time debugger only showed the location of the error. Goanna simulates this feature in the type debugging setting. Instead of run-time values, Goanna allows programmers to inspect the type assignments and observe how they change with different assumptions of the potential cause of a type error. 

    \subsection{Limitations}

    \paragraph{\textbf{Responsiveness}}
    The trade-off of extensive analysis undertaken in Goanna results in a substantial delay (mean=0.98s).
    Goanna's current performance is not yet suited for real-time feedback in programming tasks. Based on Nielsen's suggestion for waiting time tolerance \cite{Ferdowsi2023-au}, Goanna should provide responsive error diagnoses ($\leq$ 1 second) for real-time programming analysis, where users' flow of thoughts stays uninterrupted. Even in larger and more complex tasks, Goanna's response time ($\leq$ 10 seconds) is still suitable as an on-demand tool when a complex type error occurs. As shown in Wu's study \cite{Wu2017-eb}, in the simplest situation where students fix the type error in a single step, it will usually take about 60 seconds to complete the task. This error resolution time increases in proportion to the number of steps that students take to fix the error. If Goanna is able to shorten the steps to final resolution, then the querying time will easily be offset by the time it saves.

\begin{figure}[htb!]
        \centering
        \includegraphics[width=0.5\linewidth]{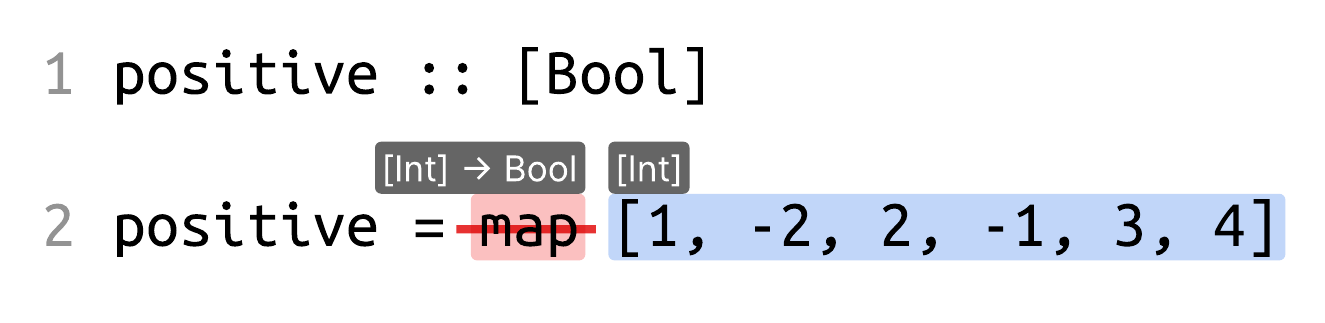}
        \caption{\textbf{} In this type error, Goanna suggests changing \texttt{map} to a function of type \texttt{[Int] -> Bool}. Although this is technically correct, in practice, a human expert user would easily identify that a function expression is missing between the function \texttt{map} and the list literal.}
        \label{fig:weakness}
    \end{figure}

    \paragraph{\textbf{Edge cases}} \label{sec:edge-case}
    Although Goanna's error reporting is generally exhaustive, there are situations where Goanna still fails to provide insightful diagnosis. One general theme is that Goanna is very effective when the error requires modifying a syntax node, but can be less insightful when the intended fix is to insert, delete, or rearrange syntax nodes. In the example in Fig.~\ref{fig:weakness}, Goanna suggests changing the \texttt{map} function to a function of type \texttt{[Int] -> [Bool]}. But in practice, a human user would very easily identify that an expression defining the function to be mapped is missing between the map and the list being operated on.
    
\subsection{Future Work}\label{sec:future-work}
  	It is important to evaluate Goanna with human participants and gain qualitative insight into its effectiveness. In one of our workshop's preliminary studies, participants showed positive reactions after using Goanna. A rigorous human study based on realistic debugging use cases is planned in the near future.

    Several areas of potential enhancement could improve Goanna's functionality and efficiency. One exciting path of improvement is to generate suggestions for syntax changes on top of type changes. As pointed out in \cite{Chen2014-dz}, syntax changes are much more challenging. But with recent improvements in generative models and the advancement of ML-based type error research \cite{Seidel2017-uf}, a precise syntax change may be on the path to becoming feasible.

    Several approaches for Goanna to achieve higher performance show promise. The parallel capability of state-of-the-art MUS enumeration \cite{Zhao2016-bu} algorithms is not explored in this study. With proper implementation, it will be possible to lower the hardware barrier of entry for a wide adoption of MCS-based type error suggestion tools. Furthermore, with domain-agnostic MUS enumeration tools \cite{Bendik2020-pz}, it should be possible to consistently achieve high performance while using a more performant constraint system and proper parameterization. Lastly, in a real-world implementation, it is possible to employ partial MUS/MCS enumeration \cite{Previti2013-mr,Liffiton2016-xi} to restrict the enumeration process in a sensible time-bound.

    Additional future work could include examining Goanna's integration with other tools within the Haskell development ecosystem. For instance, including Goanna in widely used text editors or development environments could offer developers a more integrated and fluid experience and an important avenue for Goanna to reach a wider audience. 

    Finally, it may be important to consider extending Goanna's capabilities to support other functional programming languages beyond Haskell. Several other languages, including Scala and OCaml, also support static typing and type inference, and Goanna could be a valuable addition in these contexts. In the future, we also hope to extend these techniques to popular multi-paradigm languages such as TypeScript and Rust.

    \section{Related Work} \label{sec:related-work}
	\subsection{MCS enumeration}
	
    MCS enumeration has been extensively studied in the field of error localization across various domains. In the specific context of Haskell type error diagnosis and resolution, several related approaches have been explored. It is important to examine Goanna's strengths and weaknesses in the context of these areas.

    One notable work by Lamraoui et al. introduced a tool \cite{Lamraoui2016-wr} that utilizes the ability of MCS to locate multiple faults and identify software defects using unit tests. Their approach demonstrated the effectiveness of MCS in pinpointing errors within a program. Similarly, Bekkouche et al. conducted a relevant study \cite{Bekkouche2015-is} on the utilization of MCS to locate program errors in while-loop programs. Their findings showed an improved efficiency compared to SAT-based approaches. Although showing strength in programming language static analysis, MCS-based fault localization has not been previously applied at the type system level. Goanna distinguishes itself as the first tool to explore this approach within the realm of type error diagnosis and resolution.

    \subsection{Suggesting changes to type errors}
  Lerner et al. proposed Seminal \cite{Lerner2007-mu}, using syntax mutation and binary search to find appropriate syntax changes to program errors. The advantage of Seminal is that it can suggest direct syntax changes to common mistakes (e.g., mistakingly swapping the order of function arguments). However, it is impossible for Seminal to provide the complete set of all potential fixes. Nor does it guarantee that a suggested solution is minimal syntax change.
   
   Counter-factual typing (CFT) \cite{Chen2014-dz,Chen2020-ad} uses a variation-based type system; it can suggest the correct type for all possible errors. CFT shares many capabilities with Goanna, CFT is able to suggest multiple-location changes, and CFT uses similar ranking heuristics. Goanna is able to produce an in-depth analysis of the ill-typed program, such as type error isolation. CFT and Goanna both aim to produce a complete set of potential fixes; Goanna employs a set of effective algorithms to reduce the exponential number of potential fixes without suffering the quality of suggestions. 
   
   SHErrLoc \cite{Zhang2015-xy} uses constraint-based type inference and type error diagnoses. SHErrLoc is able to suggest multiple possible fixes of the type error and rank them based on heuristics. Unlike Goanna's approach of using a well-established constraint language Prolog, SHErrLoc relies on GHC's internal constraints and then translates them into SHErrLoc Constraint Language (a custom-made constraint language). On the technical side, this approach relies heavily on modification of the compiler and is not reliable with later versions of GHC. This is indicated by the difficulties in compiling the SHErrLoc project with modern GHC versions. SHErrLoc's error diagnosis also suffers from the lack of further analysis after the error locations are identified. Goanna is able to perform type reconstruction for ill-typed programs, that is, finding the most concrete types for relevant expressions for each potential solution using the Maximal Satisfiable Subsets.

\subsection{Type Error Slicing}

Type error slicing \cite{Haack2004-fr} is a technique to identify all necessary locations of a type error that is necessary for programmers to diagnose the root cause. It has been studied in many studies since \cite{Tip2001-qn, Heeren2003-kd}. All these studies use the minimal unsatisfiable subset (MUS) to ensure the \textit{completeness} and \textit{minimality}. The drawback of type error slicing is that it often produces too many locations.  Chameleon \cite{Stuckey2003-pz,Fu2021-xd} improved type error slicing by allowing programmers to interactively show the partial MUS by choosing their own assumptions. Compared to these tools, which base their analysis on a single MUS, Goanna effectively utilizes all possible MUSes. This allows Goanna to enhance its suggestions based on improved knowledge of the underlying type of error. For example, it uses the number of MUSes a location appears in to rank the likelihood that the location is part of the root cause. This is not possible with a single MUS.  

    \section{Summary} \label{sec:conclusion}
    In this paper, we introduce Goanna, a tool to identify and resolve type errors in Haskell code. We describe the features of Goanna, including its fix suggestions, type error grouping, and identifying multiple type errors. We also discuss our approaches to reducing and prioritizing fix suggestions while maintaining comprehensiveness. Additionally, we walk through the uses of Goanna-IDE, a type error debugging interface for Haskell.

    We evaluated the effectiveness of Goanna from a set of 86 diverse Haskell programs and demonstrated its ability to identify and resolve type errors accurately compared to other tools. We also showed that Goanna effectively condenses the list of causes. When too many potential causes are present, Goanna's suggestion ranking heuristics ensures that more useful fixes are prioritized. Goanna currently works with Haskell, but in the future, we plan to extend its MCS-based error diagnosis to work with other strongly typed languages.
    
\bibliographystyle{ACM-Reference-Format}
\bibliography{paperpile}

\end{document}

%% file: preamble.tex
\AtBeginDocument{%
  }

\setcopyright{acmlicensed}
\copyrightyear{2024}
\acmYear{2024}
\acmDOI{XXXXXXX.XXXXXXX}

\acmConference[ICFP '24]{The 29th ACM SIGPLAN International Conference on Functional Programming}{September, 2024}{Milan, Italy}

\author{Shuai Fu}
\email{shuai.fu@monash.edu}
\affiliation{%
 \institution{Monash University}
 \country{Australia}
}

\author{Tim Dwyer}
\email{tim.dwyer@monash.edu}
\affiliation{%
 \institution{Monash University}
 \country{Australia}
}

\author{Peter J. Stuckey}
\email{peter.stuckey@monash.edu}
\affiliation{%
 \institution{Monash University}
 \country{Australia}
}

\author{John Grundy}
\email{john.grundy@monash.edu}
\affiliation{%
 \institution{Monash University}
 \country{Australia}
}

\begin{abstract}
Statically-typed languages offer significant advantages, such as bug prevention, enhanced code quality, and reduced maintenance costs. However, these benefits often come at the expense of a steep learning curve and a slower development pace. Haskell, known for its expressive and strict type system, poses challenges for inexperienced programmers in learning and using its type system, especially in debugging type errors. We introduce Goanna, a novel tool that serves as a type checker and an interactive type error debugging tool for Haskell. When encountering type errors, Goanna identifies a comprehensive list of potential causes and resolutions based on Minimal Correction Subset (MCS) enumeration. We evaluate Goanna's effectiveness using 86 diverse Haskell programs from online discourse, demonstrating its ability to accurately identify and resolve type errors. Additionally, we present a collection of techniques and heuristics to enhance Goanna's suggestion-based error diagnosis and show their effectiveness from our evaluation.
\end{abstract}

\ccsdesc[500]{Theory of computation~Type theory}
\ccsdesc[300]{Software and its engineering~Software testing and debugging}
\ccsdesc[300]{Software and its engineering~Fault tree analysis}
\ccsdesc[500]{Software and its engineering~Constraint and logic languages}
\ccsdesc[300]{Software and its engineering~Functional languages}
\ccsdesc[500]{Theory of computation~Constraint and logic programming}